\documentclass{ws-procs9x6-cpt22}
\begin{document}

\newcommand{\refeq}[1]{(\ref{#1})}
\def\etal {{\it et al.}}
%any other macros go here

\title{
Testing CPT Invariance by High-Precision Comparisons\\ of Fundamental Properties of Protons and Antiprotons at BASE
}

\author{E.J.\ Wursten,$^1$  M.J.\ Borchert,$^{1,2,3}$ J.A.\ Devlin,$^{1,4}$ S.R.\ Erlewein,$^{1,4,5}$ M.\ Fleck,$^{1,6}$ J.A.~Harrington,$^{1,5}$ J.I.\ J\"ager,$^{4,5}$ B.M.\ Latacz,$^{1}$ P.\ Micke,$^{1,4,5}$ G.\ Umbrazunas,$^{1,7}$  F.~Abbass,$^{8}$ M.\ Bohman,$^{{1,5}}$ S.\ Kommu,$^{{8}}$  D.\ Popper,$^{{8}}$ M.\ Wiesinger,$^{{1,5}}$ C.\ Will,$^{{5}}$ H.~Yildiz,$^{{8}}$ K.\ Blaum,$^{{5}}$ Y.\ Matsuda,$^{{6}}$ A.\ Mooser,$^{{5}}$ C.\ Ospelkaus,$^{{2,3}}$ A.\ Soter,$^{{7}}$ W.~Quint,$^{{9}}$ J.\ Walz,$^{{8}}$ Y.\ Yamazaki,$^{{1}}$ C.\ Smorra,$^{{1,8}}$ and S.\ Ulmer$^{{1}}$ }

\address{
$^1$Ulmer Fundamental Symmetries Laboratory, RIKEN, Wako, 351-0198, Japan\\
$^2$Institut f{\"u}r Quantenoptik, Leibniz Universit{\"a}t Hannover, 30167 Hannover, Germany\\
$^3$Physikalisch-Technische Bundesanstalt, 38116 Braunschweig, Germany\\
$^4$CERN, 1217 Meyrin, Switzerland\\
$^5$Max-Planck-Institut f{\"u}r Kernphysik, 69117, Heidelberg, Germany\\
$^6$Graduate School of Arts and Sciences, University of Tokyo, Tokyo 153-0041, Japan\\
$^7$Eidgenössische Technische Hochschule Zürich, 8092 Zürich, Switzerland\\
$^8$Institut f{\"u}r Physik, Johannes Gutenberg-Universit{\"a}t, 55099 Mainz, Germany\\
$^9$GSI-Helmholtzzentrum f{\"u}r Schwerionenforschung GmbH, 64291 Darmstadt, Germany
}

\begin{abstract}
The BASE collaboration at the Antiproton Decelerator facility of CERN compares the fundamental properties of protons and antiprotons using advanced Penning-trap systems.
In previous measurement campaigns, we measured the magnetic moments of the proton and the antiproton, reaching (sub-)parts-in-a-billion fractional uncertainty. 
In the latest campaign, we have compared the proton and antiproton charge-to-mass ratios with a fractional uncertainty of 16 parts in a trillion. 
In this contribution, we give an overview of the measurement campaign, and detail how its results are used to constrain nine spin-independent coefficients of the Standard-Model Extension in the proton and electron sector.
\end{abstract}

\bodymatter

%%%%%%%%%%%%%%%%%%%%%%%%%%%%%%%%%
\section{Introduction} 
%%%%%%%%%%%%%%%%%%%%%%%%%%%%%%%%%

The Baryon Antibaryon Symmetry Experiment (BASE) collaboration is dedicated to the study of protons and antiprotons with ultrahigh precision.
Comparisons of fundamental properties of such matter--antimatter conjugates provide tests of CPT invariance, one of the cornerstones of the Standard Model of Particle Physics.
The BASE collaboration operates several Penning-trap experiments at different locations.
At CERN's Antiproton Decelerator facility, primarily two types of measurements take place.
Firstly, the charge-to-mass ratio of the antiproton is compared (in situ) to that of the negatively-charged hydrogen ion, currently reaching a fractional uncertainty of 16 parts per trillion.\cite{Borchert2022}
Secondly, the magnetic moment of the antiproton is measured. 
The magnetic moment of the proton is measured at a dedicated experiment in Mainz,\cite{Schneider2017} reaching a fractional uncertainty of 300 parts per trillion, which is 5 times more precise than the antiproton experiment.\cite{Smorra2017}
These comparisons of fundamental properties can be interpreted in terms of the Standard-Model Extension (SME),\cite{datatables} yielding constraints on different types of coefficients of the model in the proton (and electron) sector.

%%%%%%%%%%%%%%%%%%%%%%%%%%%%%%%%%
\section{Comparing antiproton and proton charge-to-mass ratios} 
%%%%%%%%%%%%%%%%%%%%%%%%%%%%%%%%%

The principle of an antiproton-to-proton charge-to-mass ratio comparison at BASE is the measurement of the cyclotron frequency $\omega_{\mathrm{c}}{=}q B/m$  of the two particles while stored in a Penning trap.
Here, $B$ is the magnetic field of the Penning trap and $(q/m)$ the charge-to-mass ratio of the trapped particle.
The cyclotron frequency $\omega_{\mathrm{c}}$ is inferred from the three eigenmotions of the trapped particle, which are detected by monitoring the image currents induced in the electrodes of the Penning trap.

To suppress systematic effects related to residual magnetic-field gradients and trap-polarity inversion,\cite{Ulmer2015} the antiproton is not directly compared to the proton, but to the negatively charged hydrogen ion H$^-$.
The mass of the H$^-$ ion is given by
\begin{equation}
m_{\text{H}^-}=m_{\mathrm{p}}\left(1 + 2\frac{m_e}{m_{\mathrm{p}}} - \frac{E_\mathrm{b}}{m_{\mathrm{p}}c^2} -\frac{E_\mathrm{a}}{m_{\mathrm{p}}c^2} +\frac{\alpha_{\text{H}^-} B^2}{m_{\mathrm{p}}c^2}\right) = m_{\mathrm{p}} R.
\end{equation}
Here, $m_e$ is the electron mass, $E_\mathrm{b}$ is the binding energy of the electron in hydrogen, $E_\mathrm{a}$ is the afﬁnity energy of the second electron, and the last term accounts for the electrical polarizability $\alpha_{\text{H}^-}$ of the hydrogen ion.
Measurements of the cyclotron frequencies of the antiproton $\bar{\text{p}}$ and the H$^-$ ion in the same magnetic field $B$ give access to the dimensionless ratio of charge-to-mass ratios
\begin{equation}
R_{\bar{\mathrm{p}},\mathrm{H}^-}=\frac{\omega_{\mathrm{c},\bar{\mathrm{p}}}}{\omega_{\mathrm{c},\mathrm{H}^-}}= \frac{(q/m)_{\bar{\mathrm{p}}}}{(q/m)_{\mathrm{H}^-}}\,.
\end{equation}
Assuming that CPT invariance holds, the theoretical prediction is that $R_{\bar{\mathrm{p}},{\text{H}^-}}$ is identical to 
$R=1{.}001\,089\,218\,753\,80(3)$.\cite{Ali2020}

%%%%
\begin{figure}[htb]
\begin{center}
\includegraphics[width=\textwidth,keepaspectratio]{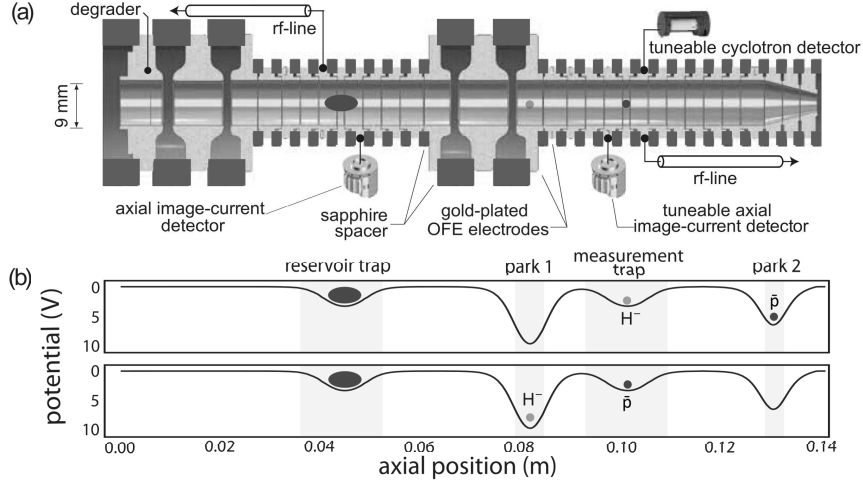}
\end{center}
\caption{(a) Illustration of the BASE multi-Penning-trap system. (b) Configurations of the Penning-trap potentials for cyclotron-frequency measurements of H$^-$ (top) and $\bar{\text{p}}$ (bottom).}
\label{fig:BASEtrap}
\end{figure}
In practice, the cyclotron frequency-measurements on $\bar{\text{p}}$ and H$^-$ are not done simultaneously in the same Penning trap, but one after the other, shuttling the particles into and out of a dedicated measurement trap.
A sketch of the multi-Penning-trap system that is used in these experiments is shown in Fig.\ \ref{fig:BASEtrap}(a).
A cloud of H$^-$ and $\bar{\text{p}}$ delivered by the Antiproton Decelerator is stored in the reservoir trap on the left.
A single ion of each species is extracted from the reservoir trap and can be stored in different park electrodes along the trap stack on the right or shuttled into the measurement trap for a cyclotron-frequency measurement.
Figure~\ref{fig:BASEtrap}(b) shows the Penning-trap-potential configurations that are used to perform cyclotron-frequency measurements on H$^-$ (top) and $\bar{\text{p}}$ (bottom).
The measurement of one individual cyclotron frequency takes about $110\,$s. 
Together with the required particle exchange, the determination of a single frequency ratio $R_{\bar{\mathrm{p}},\mathrm{H}^-}$ takes about 4$\,$min. 
We have accumulated a total of about 24\,000 such cyclotron-frequency-ratio measurements using different techniques during the last measurement campaign, spanning 1.5$\,$years of sampling time.
The resulting ratio of antiproton-to-proton charge-to-mass ratios is
\begin{equation}
\frac{(q/m)_{\bar{\mathrm{p}}}}{(q/m)_{\mathrm{p}}} = R_{\bar{\text{p}},\text{H}^-} /R=-1.000\,000\,000\,003\, (16)\,,\cite{Borchert2022}
\label{eq:QMresult}
\end{equation}
constituting an improvement of a factor of 4 compared to the previous measurement campaign.\cite{Ulmer2015}
This improvement was possible due to several upgrades of the apparatus, eliminating previous systematic limitations and considerably improving the stability of the experiment.

%%%%%%%%%%%%%%%%%%%%%%%%%%%%%%%%%
\section{Constraining SME coefficients with BASE} 
%%%%%%%%%%%%%%%%%%%%%%%%%%%%%%%%%
The results of a charge-to-mass ratio comparison can be interpreted as a test of Lorentz and CPT symmetry, as outlined in detail in Ref.\ [\refcite{Ding2020}].
Applying the SME framework to the Penning trap, the energy levels $E_{n,s}^w$ and corresponding energy shifts ${\delta}E_{n,s}^w$ due to the SME terms in the Lagrangian can be calculated for a trapped particle.
Here, $n=0,1,2,...$ indicates the relativistic Landau level number, $s=\pm1$ the spin state relative to the magnetic field $B$ and $w$ the species of the particle. 
The cyclotron frequencies that we measure in our experiment can be expressed as differences between energy levels $\hbar \omega_{\mathrm{c},w}^s = E_{1,s}^w-E_{0,s}^w$, giving the dependence of the measured cyclotron frequencies on the SME coefficients, as laid out in Ref.\ [\refcite{Ding2020}].
Do note that we distinguish the spin state~$s$ in our definition of the cyclotron frequency $\omega_{\mathrm{c},w}^s$, as both spin orientations are possible during our frequency measurements in the Penning trap.
The shift of the antiproton cyclotron frequency is then written as $\delta\omega_{\mathrm{c},\bar{\mathrm{p}}}^{s} = {\delta}E_{1,s}^{\bar{\mathrm{p}}}-{\delta}E_{0,s}^{\bar{\mathrm{p}}} $, whereas the shift of the $\mathrm{H}^-$ frequency can be approximated as
\begin{equation}
\delta\omega_{\mathrm{c},\mathrm{H}^-} = \delta\omega_{\mathrm{c},\mathrm{p}}^{s} + \left(\delta\omega_{\mathrm{c},e}^{\uparrow} +\delta\omega_{\mathrm{c},e}^{\downarrow}\right),
\end{equation}
where the last term takes into account that the electron pair is in a singlet spin state.
The charge-to-mass ratio comparison of Eq.\ \refeq{eq:QMresult} can then be interpreted as follows:
\begin{equation}
\frac{(|q|/m)_{\bar{\mathrm{p}}}}{(|q|/m)_{\mathrm{p}}}-1  = \frac{\omega_{\mathrm{c},\bar{\mathrm{p}}}}{R\,\omega_{\mathrm{c},\mathrm{H}^-}}-1
= \frac{\delta\omega_{\mathrm{c},\bar{\mathrm{p}}}^{s1} - R\,\delta\omega_{\mathrm{c},\mathrm{p}}^{s2} - R\,\left(\delta\omega_{\mathrm{c},e}^{\uparrow} +\delta\omega_{\mathrm{c},e}^{\downarrow}\right)}{R\,\omega_{\mathrm{c},\mathrm{H}^-}},
\label{eq:RatioSME}
\end{equation}
yielding the following limit for the time-averaged component of the model
\begin{equation}
\hbar|\delta\omega_{\mathrm{c},\bar{\mathrm{p}}}^{s1} - R\,\delta\omega_{\mathrm{c},\mathrm{p}}^{s2} - R\,\left(\delta\omega_{\mathrm{c},e}^{\uparrow} +\delta\omega_{\mathrm{c},e}^{\downarrow}\right)| < 1.96\times 10^{-27}\,\mathrm{GeV}.
\label{eq:BASE_SMEconstraint}
\end{equation}
After transforming from the local laboratory frame to the standard Sun-centered inertial reference frame,\cite{datatables} we can use Eq.\ \refeq{eq:BASE_SMEconstraint} to individually constrain the nine spin-independent SME coefficients summarized in 
Table~\ref{tab:BASE_SMEcoeff}.
\begin{table*}[h!]
\tbl{Individual constraints on spin-independent electron and proton coefficients of the SME.}
{\begin{tabular}{l c } \toprule
Coefficient  &  Limit (68\% C.L.)  \\
\colrule
$|\tilde{c}_e^{XX}|$, $|\tilde{c}_e^{YY}|$ & $<7.79\cdot10^{-15}$  \\
$|\tilde{c}_e^{ZZ}|$ & $<4.96\cdot10^{-15}$  \\
\colrule
$|\tilde{c}_{\mathrm{p}}^{XX}|, |\tilde{c}_{\mathrm{p}}^{*XX}|$, $|\tilde{c}_{\mathrm{p}}^{YY}|, |\tilde{c}_{\mathrm{p}}^{*YY}|$  & $<2.86\cdot10^{-11}$  \\
$|\tilde{c}_{\mathrm{p}}^{ZZ}|, |\tilde{c}_{\mathrm{p}}^{*ZZ}|$  &  $<1.82\cdot10^{-11}$   \\
[1ex] 
\botrule
\end{tabular}
}
\label{tab:BASE_SMEcoeff}
\end{table*}
Since the spin state of the used antiprotons and hydrogen ions is not known a priori nor measured, we cannot constrain the spin-dependent coefficients contained within $\delta\omega_{\mathrm{c},\bar{\mathrm{p}}}^{s}$ and $\delta\omega_{\mathrm{c},\mathrm{p}}^{s}$.
Similarly, due to the singlet state of the electron pair, all spin-dependent electron coefficients cancel out and cannot be constrained either.

%%%%%%%%%%%%%%%%%%%%%%%%%%%%%%%%%
\section{Outlook} 
%%%%%%%%%%%%%%%%%%%%%%%%%%%%%%%%%
We can study the temporal variations of the ratio of antiproton-to-proton charge-to-mass ratios at harmonics of the sidereal day and the sidereal year as we have a sampling period of 4\,min and taken data for 1.5\,years.
Such an analysis is currently ongoing and will significantly expand the number of SME coefficients that can be constrained.

In addition, we are currently working towards a ten-fold improved measurement of the antiproton magnetic moment, which will allow us to improve constraints on another twelve coefficients of the SME.
Furthermore, there are plans to integrate the spin-detection techniques developed for the current magnetic-moment campaign into the next charge-to-mass ratio campaign. Knowing the antiproton and H$^-$ spin state would give us access to the spin-dependent proton coefficients listed in Ref.\ [\refcite{Ding2020}].

%%%%%%%%%%%%%%%%%%%%%%%%%%%%%%%%%
\section*{Acknowledgments} 
%%%%%%%%%%%%%%%%%%%%%%%%%%%%%%%%%

We acknowledge technical support by CERN. We acknowledge financial support by the RIKEN EEE pioneering project funding, the RIKEN SPDR and JRA program, the Max Planck Society, the European Union (FunI-832848,  STEP-852818), CRC 1227 ``DQ-mat"(DFG 274200144), the Cluster of Excellence ``Quantum Frontiers" (DFG 390837967), AVA-721559, the CERN fellowship program, and the Helmholtz-Gemeinschaft. This work was supported by the Max-Planck--RIKEN--PTB Center for Time, Constants, and Fundamental Symmetries (C-TCFS).

\end{document}